\documentclass[twocolumn,showpacs,floats,floatfix,superscriptaddress,aps,pra]{revtex4-1}
\usepackage{amsfonts}
\usepackage{amssymb}
\usepackage{amsmath}
\usepackage{calc}
\usepackage{graphicx}
\usepackage{bm}

\usepackage[normalem]{ulem}
\usepackage{amsmath,amssymb}
\usepackage{multirow}
\usepackage{xcolor,soul}
\usepackage{srcltx}
\usepackage{hyperref,graphicx}

\def\be{ \begin{equation}}
\def\ee{ \end{equation}}
\def\bea{ \begin{eqnarray}}
\def\eea{ \end{eqnarray}}
\def\bse{ \begin{subequations}}
\def\ese{ \end{subequations}}
\def\bc{ \begin{center}}
\def\ec{ \end{center}}

\begin{document}

\author{Stefano Longhi$^{*}$} 
\affiliation{Dipartimento di Fisica, Politecnico di Milano and Istituto di Fotonica e Nanotecnologie del Consiglio Nazionale delle Ricerche, Piazza L. da Vinci 32, I-20133 Milano, Italy}
\email{stefano.longhi@polimi.it}

\title{Non-Hermitian bidirectional robust transport}
  \normalsize


%
\bigskip
\begin{abstract}
\noindent  

Transport of quantum or classical waves in open systems is known to be strongly affected by non-Hermitian terms that arise from an effective description of system-enviroment interaction. A simple and paradigmatic example of non-Hermitian transport, originally introduced by Hatano and Nelson two decades ago [N. Hatano and D.R. Nelson, Phys. Rev. Lett. {\bf 77}, 570 (1996)], is  the hopping dynamics of a quantum particle on a one-dimensional tight-binding lattice in the presence of an {\em imaginary} vectorial potential. The imaginary gauge field can prevent Anderson localization via non-Hermitian delocalization, opening up a mobility region and realizing robust transport immune to disorder and backscattering. Like for robust transport of topologically-protected edge states in quantum Hall and topological insulator systems, non-Hermitian robust transport in the Hatano-Nelson model is {\it unidirectional}. However, there is not any physical impediment to observe robust {\it bidirectional}  non-Hermitian transport. Here it is shown that in a quasi-one dimensional zigzag lattice, with non-Hermitian (imaginary) hopping amplitudes and a synthetic  gauge field, robust transport immune to backscattering can occur bidirectionally along the lattice.

\end{abstract}



\maketitle

\section{Introduction}
Transport, localization and scattering of quantum or classical waves in systems described by effective non-Hermitian Hamiltonians are of major interest in different areas of science \cite{r1,r2,r3,r4,r5,r6,r7,r8,r9,r10,r11,r13,r14,r15,r16,r17,r18,r19,r20,r21,r22,r23,r24,r25,r26,r27,r28,r29,r30,r30bis,r31,r32,r33,r34,r35,r36,r37,r38,r39,r40,r41,r42,r42bis,r43,r43bis,r44,r45,r46,r47,r48,r49,r50}, ranging from the physics of open quantum systems to mesoscopic solid-state structures \cite{r2,r14,r15,r18,r21,r24,r26}, atomic and molecular physics \cite{r1,r34}, optics and photonics \cite{r19,r30,r31,r36,r37,r39,r40,r42bis}, acoustics \cite{r46,r50}, magnetic and spin systems \cite{r22,r29,r32,r33,r35,r44,r45}, quantum computing \cite{r27,r28,r42}, and biological systems \cite{r7,r48}. Several important signatures of non-Hermitian transport have been revealed, including non-Hermitian delocalization in disordered lattices \cite{r3,r4,r5,r6,r7,r8,r9,r10,r11,r13}, one-way scattering \cite{r16,r17}, transition from ballistic to diffusive transport \cite{r30}, hyperballistic transport \cite{r30bis}, invisibility of defects \cite{r23}, topological phase transitions \cite{r20,r24,r39,r40}, and non-Hermitian transparency \cite{r38} to mention a few. 

A simple and paradigmatic model of non-Hermitian transport 
was introduced in 1996 by Hatano and Nelson \cite{r3}, who investigated the problem of Anderson localization in a one-dimensional disordered non-Hermitian lattice and predicted the existence of so-called {\it non-Hermitian} delocalization transition \cite{r4,r5,r6,r7,r8,r9,r10,r11,r13}. They showed that an 'imaginary' vector field, leading to an imaginary Peierls' phase in the hopping amplitude, can prevent Anderson localization owing to the appearance of a mobility interval at the center of the band. Inspired by the non-Hermitian delocalization transition and the possibility to get mobility regions in one-dimensional lattices in spite of disorder, in a recent work \cite{r37} it has been conjectured that robust transport, which is rather insensitive to defects or lattice disorder, should arise in a rather general class of non-Hermitian one-dimensional lattices with properly-tailored energy dispersion band (see also \cite{r38}). Non-Hermitian robust transport, such as the one found in the Hatano-Nelson model, turns out to be {\it unidirectional} \cite{r37}, i.e. it is observed only for waves propagating along one direction of the lattice, but not in the opposite direction. Such a unidirectional  property is in some sense similar to the directiveness (chirality)  of edge states in robust transport at the boundary of two- or three-dimensional  quantum Hall systems, topological insulators, and
topological superconductors (see, for instance, \cite{r51,r52,r53,r54} and references therein), although in the non-Hermitian case robustness of transport does not rely on topological protection. However, there is not any basic physical reason that prevents non-Hermitian transport to work bidirectionally. \par
In this work we show that {\it bidirectional} non-Hermitian robust transport can be observed in quasi one-dimensional tight-binding lattices. Specifically, we consider a zigzag lattice with non-Hermitian (complex) hopping amplitudes in the presence of a synthetic (real) gauge field, and show that robust transport, immune to backscattering, can be observed for both forward and backward propagating waves along the lattice for proper tuning of the synthetic gauge field.

\section{Non-Hermitian zigzag lattice model}
We consider quantum or classical transport in the quasi one-dimensional non-Hermitian zigzag lattice schematically shown in Fig.1(a). The lattice comprises two sublattices A and B with an imaginary (non-Hermitian) hopping amplitude $-i \kappa$ between adjacent sites in each sublattice and with an Hermitian inter-site hopping rate $\rho$ with a phase controlled by a synthetic gauge field.
 Indicating by $a_n$ and $b_n$ the occupation amplitudes of the sites $|n \rangle_A$ and $|n \rangle_B$  in the two sublattices A and B, the evolution equations for $a_n$ and $b_n$ read
 \begin{subequations}
 \begin{eqnarray}
 i \frac{da_n}{dt} & = & (-i \gamma+V^{(A)}_n)a_n-i \kappa(a_{n+1}+  a_{n-1}) \nonumber \\
 & + &  \rho(\exp(i \varphi) b_n+ \exp(i \varphi')b_{n-1})\\
 i \frac{db_n}{dt} & = & (-i \gamma+V^{(B)}_n)b_n-i \kappa(b_{n+1}+  b_{n-1}) \nonumber \\
 & + &  \rho(\exp(-i \varphi) a_n+ \exp(-i \varphi')a_{n+1}).
 \end{eqnarray}
 \end{subequations}
 where $\varphi$ and $\varphi'$ are the magnetic fluxes (Peierls phases)
 in the lower and upper triangular plaquettes, $V_n^{(\alpha)}$ is the energy potential at site $n$ that accounts for lattice defects or disorder, and $\gamma>0$ is an effective loss  rate. For a purely dissipative system the constraint $\gamma \geq 2 \kappa$ should be satisfied.  Note that the limit $\gamma=\kappa=0$ corresponds to an Hermitian lattice.  An effective imaginary hopping rate $-i\kappa$ can be realized by non-Hermitian engineering using auxiliary lossy sites in the geometrical setting shown in Fig.1(b), as discussed in Ref.\cite{r55}. In fact, indicating by $A_n$ and $B_n$ the occupation amplitudes in auxiliary lossy sites in the upper and lower sublattices, the coupled equations for the zigzag lattice of Fig.1(b) read 
 \begin{subequations}
  \begin{eqnarray}
 i \frac{da_n}{dt} & = & (-i \gamma+V^{(A)}_n)a_n+ \epsilon (a_{n+1}+  a_{n-1}) \nonumber \\
 & + &  \rho(\exp(i \varphi) b_n+ \exp(i \varphi')b_{n-1}) \nonumber \\
 & + & \sigma (A_{n}+A_{n+1}) \\
 i \frac{db_n}{dt} & = & (-i \gamma+V^{(B)}_n)b_n+ \epsilon (b_{n+1}+  b_{n-1}) \nonumber \\
 & + &  \rho(\exp(-i \varphi) a_n+ \exp(-i \varphi')a_{n+1}) \nonumber \\
 & + & \sigma (B_n+B_{n+1}) \\
 i \frac{dA_n}{dt} & = & U A_n+ \sigma( a_{n}+a_{n-1}) \\
 i \frac{dB_n}{dt} & = & U B_n+ \sigma( b_{n}+b_{n-1}) 
  \end{eqnarray}
 \end{subequations}
 where  $\sigma$ and $\epsilon$ are the (Hermitian) nearest neighbor and the next-to-nearest hopping  amplitudes, respectively, in the two sublattices and $U$ is the site energy of the lossy sites, with ${\rm Im}(U)<0$ [Fig.1(b)]. For $|U|$ large as compared to $\sigma$, one can eliminate adiabatically the auxiliary site amplitudes  $A_n$ and $B_n$, i.e.
 \begin{equation}
 A_n \simeq -\frac{\sigma}{U} (a_n+a_{n-1}) \; ,\; \; B_n \simeq -\frac{\sigma}{U} (b_n+b_{n-1}).
 \end{equation}
 as one readily obtains by letting $dA_n/dt \simeq 0$ and $dB_n/dt \simeq 0$ in Eqs.2(c) and 2(d).
 Substitution of Eq.(3) into Eqs.(2a) and (2b) yields
 \begin{subequations}
  \begin{eqnarray}
 i \frac{da_n}{dt} & = & \left( -i \gamma+V^{(A)}_n+\delta \right) a_n+ \kappa_{eff} (a_{n+1}+  a_{n-1}) \nonumber \\
 & + &  \rho(\exp(i \varphi) b_n+ \exp(i \varphi')b_{n-1}) \\
 i \frac{db_n}{dt} & = & \left(-i \gamma+V^{(B)}_n +1\delta \right)b_n+ \kappa_{eff} (b_{n+1}+  b_{n-1}) \nonumber \\
 & + &  \rho(\exp(-i \varphi) a_n+ \exp(-i \varphi')a_{n+1}) 
  \end{eqnarray}
 \end{subequations}
 where we have set $\kappa_{eff}= \epsilon-\sigma^2/U$ and $\delta=-2 \sigma^2/U$. Note that the additional term $\delta$ in Eqs.(4a) and (4b) just corresponds to a shift of the site complex 
 energy $\gamma$, whereas $\kappa_{eff}$ gives an effective non-Hermitian hopping amplitude between adjacent sites in the two sublattices. To achieve $\kappa_{eff}=-i \kappa$ with $\kappa$ real, thus realizing the zigzag model of Fig.1(a), it is enough to tailor the site energy $U$ of the auxiliary sites at the value $U= \sigma^2 (\epsilon-i \kappa)/(\epsilon^2 + \kappa^2)$. Correspondingly, the energy shift in the main lattice sites is $\delta=-2 \epsilon-2 i \kappa$.\\
 Let us consider the case of a lattice without site energy defects or disorder, i.e. $V_n^{(A)}=V_n^{(B)}=0$. The dispersion curves $E=E_{\pm}(q)$ of the two mini bands sustained by the zigzag lattice can be readily found by looking for a solution to Eqs.(1a) and (1b) of the form
 \begin{equation}
 \left(
 \begin{array}{c}
 a_n \\
 b_n
 \end{array}
 \right)=\left(
 \begin{array}{c}
A \\
B
 \end{array}
 \right) \exp(2iqn-iEt)
 \end{equation}
 where $-\pi/2 \leq q  \leq \pi/2$ is the Bloch wave number and $E=E(q)$ is the energy dispersion curve. One obtains
 \begin{equation}
 E_{\pm}(q)=- i \gamma - 2 i \kappa \cos (2 q) \pm 2 \rho \cos (q+ \Delta \varphi)
 \end{equation}
 where we have set $\Delta \varphi=(\varphi-\varphi')/2$. Equation (6) shows that the mini band dispersion curves depend solely on the difference $\Delta \varphi$ between the magnetic fluxes in the lower and upper plaquettes, so that without loss of generality we can assume $\varphi'=-\varphi$ and $\Delta \varphi= \varphi$. For the following analysis, it is worth introducing  a linear one-dimensional lattice with long-range hopping, which turns out to be equivalent to the zigzag lattice, as shown in Fig.1(c). In fact, after setting $c_n=a_{n/2}$, $U_n=V_{n/2}^{(A)}$ for $n$ even and  $c_n=b_{(n-1)/2}$, $U_n=V_{(n-1)/2}^{(B)}$ for $n$ odd, from Eqs.(1a) and (1b) one obtains
 \begin{eqnarray}
 i \frac{dc_n}{dt} & = & (-i \gamma+i U_n)c_n-i\kappa(c_{n+2}+c_{n-2}) \nonumber \\
 &+ & \rho(\exp(i \varphi) c_{n+1}+\exp(-i \varphi) c_{n-1})).
 \end{eqnarray}
 Such coupled equations describe the dynamics of the one-dimensional lattice with nearest- and next-nearest-neighbor hopping shown in Fig.1(c). For an ordered lattice $(U_n=0$), the dispersion curve of the linear lattice Eq.(7), obtained by setting $c_n \sim \exp(iqn-iEt)$, reads
 \begin{equation}
 E(q)=-i\gamma-2i \kappa \cos(2q)+2 \rho \cos(q+ \varphi).
 \end{equation}
 \begin{figure}[htbp]
  \includegraphics[width=86mm]{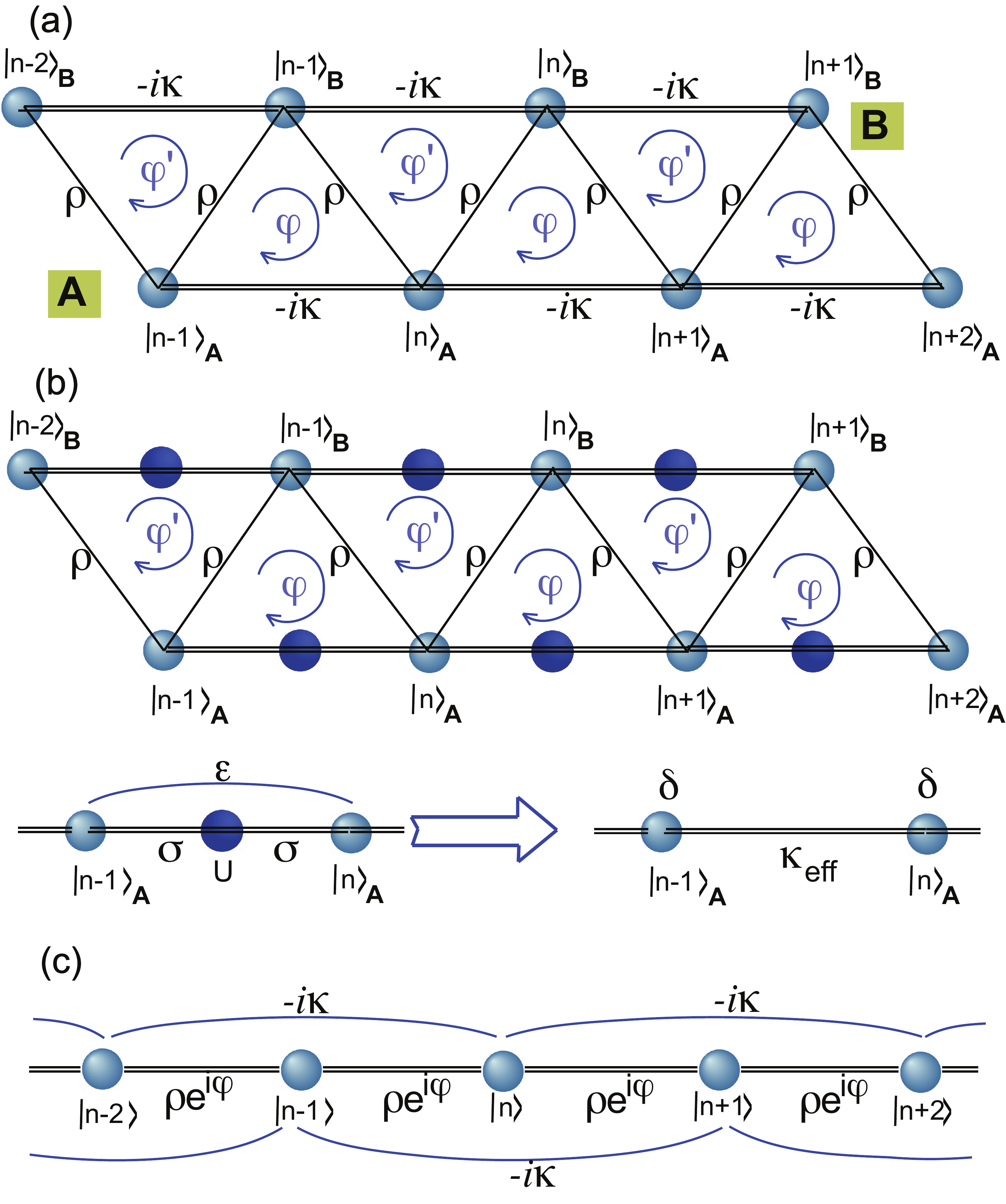}\\
   \caption{(color online) (a) Schematic of the non-Hermitian quasi one-dimensional  zigzag lattice. (b) Physical realization of the lattice with effective non-Hermitian hopping rate $-i \kappa$ obtained using auxiliary lossy sites (dark circles). Adiabatic elimination of the auxiliary sites yields an effective hopping rate $\kappa_{eff}= \epsilon-\sigma^2/U$ and (complex) site energy shift $\delta=-2 \sigma^2/U$ in the main lattice sites. (c) Equivalent linear tight-binding lattice with nearest and next-to-nearest neighbor hopping.  }
\end{figure}
 \begin{figure}[htbp]
  \includegraphics[width=86mm]{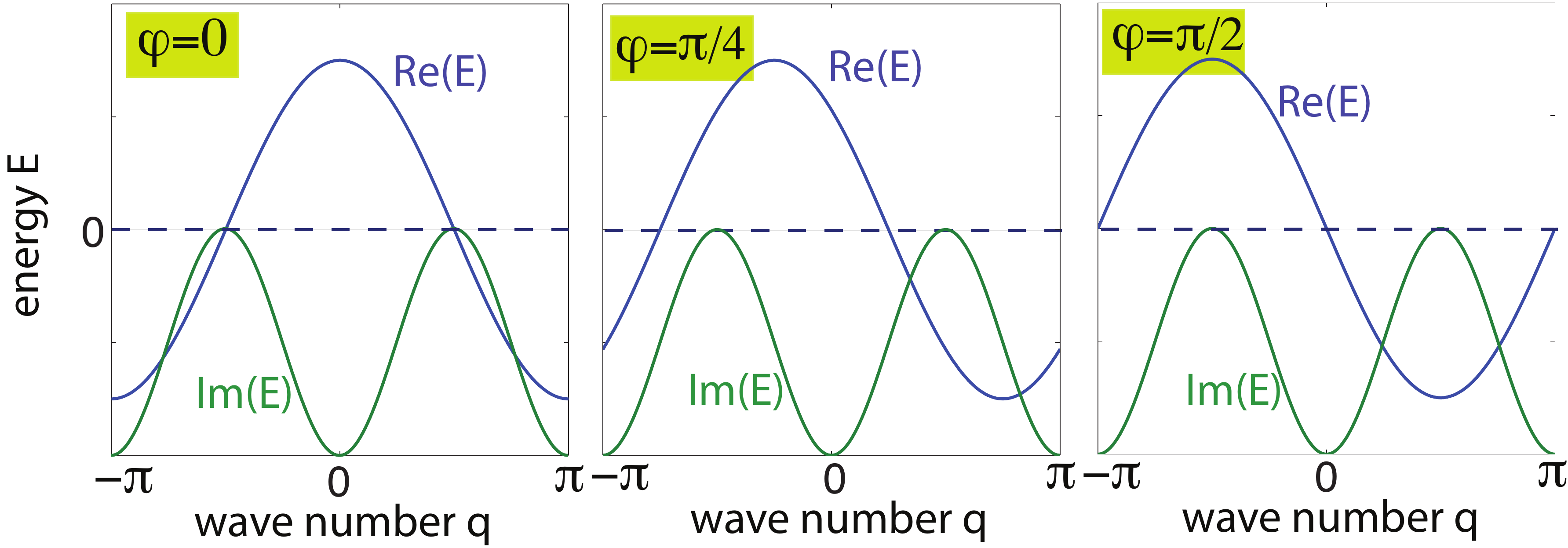}\\
   \caption{(color online) Energy band diagram of the zigzag lattice in the extended Brillouin zone [Eq.(8)] for a few values of the flux $\varphi$ and for $\gamma=2 \kappa$. The energy scale is in arbitrary units.}
\end{figure}
  Note that, since $q$ now varies in the extended Brillouin zone $-\pi \leq q \leq \pi$, Eq.(8) reproduces the two mini bands $E_{\pm}(q)$ [Eq.(6)] of the zigzag lattice. Typical behaviors of the energy dispersion curve $E(q)$ are shown in Fig.2 for non-Hermitian lattices with imaginary hopping $-i\kappa$, loss rate $\gamma= 2 \kappa$ and for the three different values of the flux $\varphi=0, \pi/4, \pi/2$. Note that a change of the flux $\varphi$ corresponds to a shift of the real part of the energy band, whereas the imaginary part of energy band is not changed. The maximum value of ${\rm Im}(E)$ is attained at the wave numbers $q_{1,2}= \pm \pi/2$. Note that for $\varphi=0$ one has $E(q_1)=E(q_2)$. 
  
  \section{Single-site excitation dynamics and non-Hermitian delocalization}
  Let us consider the spreading dynamics corresponding to initial single-site excitation of the lattice, i.e. to the initial condition $c_n(0)=\delta_{n,0}$. Let us first consider the case of an ordered lattice, i.e. $U_n=0$, and let us assume for the sake of definiteness a passive system with a loss rate $\gamma=2\kappa$; note that a change of $\gamma$ just results in an exponential change (growth or decay) of the occupation amplitudes $c_n(t)$, as it follows from Eq.(7) after the 'imaginary' gauge transformation $c_n(t) \rightarrow c_n(t) \exp(-\gamma  t)$. For $U_n=0$ the solution $c_n(t)$ to Eq.(7) with the initial condition $c_n(0)=\delta_{n,0}$ can be written formally as
  \begin{equation}
  c_n(t)=\frac{1}{2 \pi} \int_{-\pi}^{\pi} dq \exp [iqn-iE(q)t]
  \end{equation}
  where $E(q)$ is the dispersion relation defined by Eq.(8). Examples of spreading dynamics for the three values  $\varphi=0, \pi/4, \pi/2$ of the flux are shown in Fig.3.
In the figure the spreading dynamics in the Hermitian limit $\gamma=\kappa=0$ is also shown for comparison. We are interested to study the asymptotic behavior of $|c_n(t)|^2$ as $t \rightarrow \infty$, which determines the asymptotic spreading law in the lattice. Such a behavior can be readily obtained by use of the steepest descend method. For long times $t$ and $\kappa>0$, the main contributions to the integral on the right hand side of Eq.(9) comes from wave numbers $q$ around the two points $q=q_1=\pi/2$ and $q=q_2=-\pi/2$, where the imaginary part of the dispersion curve $E(q)$ reaches it maximum value. After expanding $E(q)$ at around $q=q_1$ and $q=q_2$ up to second order, for $t \rightarrow \infty$ one has
  \begin{eqnarray}
  c_n(t) & \sim  & \frac{\exp(iq_1 n-iE_1 t)}{2 \pi} \int d \xi \exp \left[ i \xi(n-E_1't)-iE_1''t \xi^2 \right] \nonumber \\
  & + & \frac{\exp(iq_2 n-iE_2 t)}{2 \pi} \int d \xi \exp \left[ i \xi(n-E_2't)-iE_2'' t \xi^2 \right] \nonumber \\
  & & 
  \end{eqnarray}
   \begin{figure}[htbp]
  \includegraphics[width=86mm]{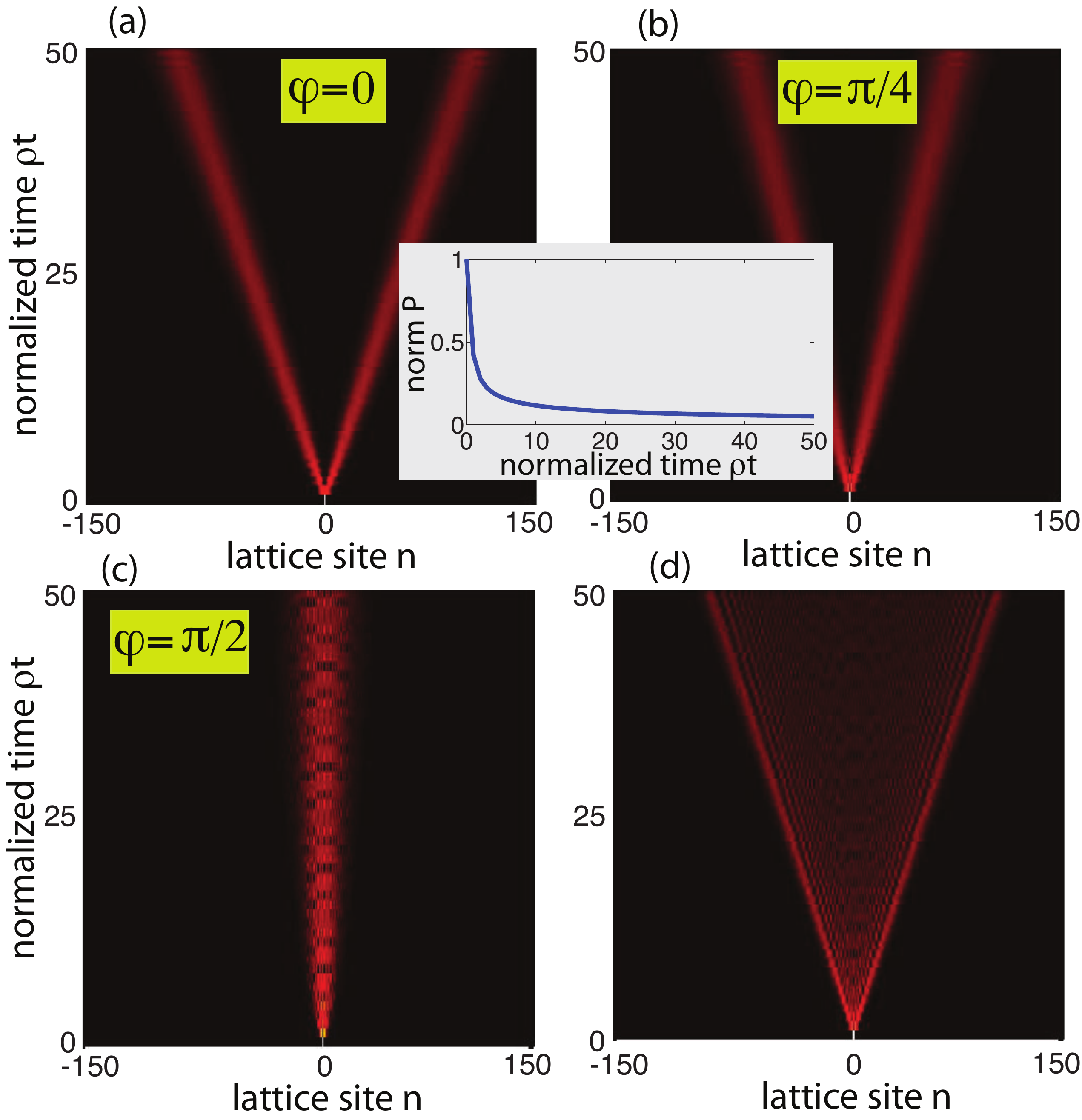}\\
   \caption{(color online) Spreading dynamics in an ordered lattice corresponding to single-site excitation at initial time $t=0$. The figures show on a pseudo color map the evolution versus time $t$  of the normalized amplitude probability distribution $|p_n(t)|$, where $p_n(t)=c_n(t)/ \sqrt{P(t)}$ and $P(t)=\sum_n |c_n(t)|^2$ is the norm. (a-c) Behavior of $|p_n(t)|$ in the non-Hermitian lattice for parameter values $\kappa=0.3$, $\rho=1$, $\gamma=2 \kappa=0.6$  and for a few values of the flux $\varphi$. The inset in the figures shows the evolution of the norm $P(t)$, which is independent of $\varphi$. (d) Behavior of $|p_n(t)|$ in the limit of an Hermitian lattice  ($\kappa=\gamma=0$, $\rho=1$). In the Hermitian limit the spreading pattern does not depend on the flux $\varphi$. }
\end{figure}
   \begin{figure}[htbp]
  \includegraphics[width=86mm]{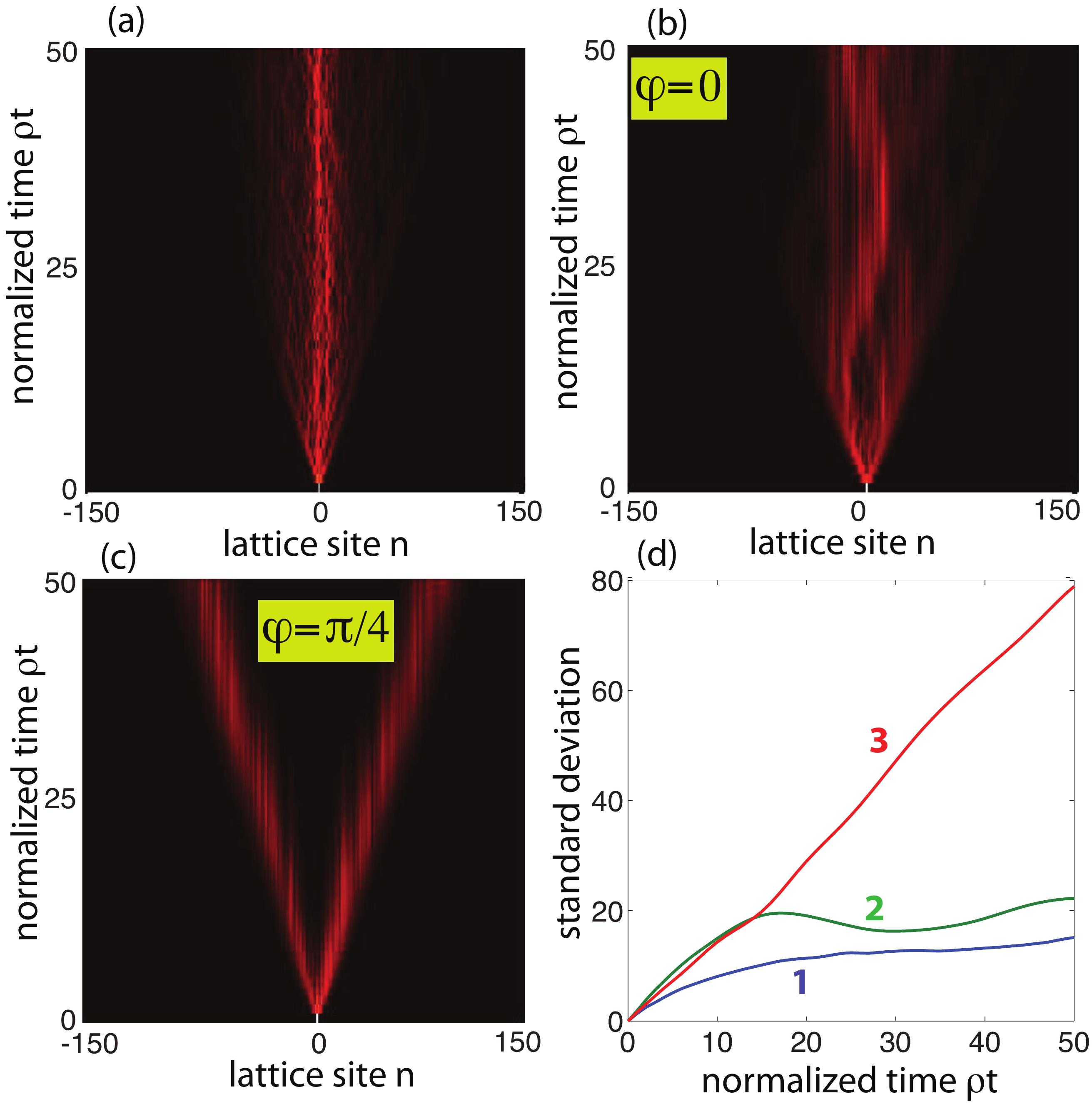}\\
   \caption{(color online) Spreading dynamics in a lattice with on-site energy disorder $U_n$ and single-site excitation at initial time $t=0$. (a-c) Evolution versus time $t$  of the normalized amplitude probability distribution $|p_n(t)|$ on a pseudo color map in the Hermitian lattice [panel (a), $\kappa=\gamma=0$, $\rho=1$, $\varphi=0$] and in the non-Hermitian lattice  ($\kappa=0.3$, $\rho=1$, $\gamma=2 \kappa=0.6$)  for $\varphi=0$ [panel (b)] and $\varphi= \pi/4$ [panel (c)]. $U_n$ is assumed to be a random real variable with uniform distribution in the range $(-\Delta, \Delta)$, with $\Delta=1$. (d) Temporal behavior of the standard deviation of the probability distribution $|p_n(t)|^2$ for the spreading patterns shown in (a), (b) and (c). Curve 1 corresponds to (a), curve 2 to (b) and curve 3 to (c). Anderson localization is observed in (a) and (b), whereas delocalization persists despite disorder in (c).}
\end{figure}
  where we have set $E_1=E(q_1)=- 2 \rho \sin \varphi$, $E_2=E(q_2)=2 \rho \sin \varphi$, $E_1'=(dE/dq)_{q_1}=- 2 \rho \cos \varphi$, $E_2'=(dE/dq)_{q_2}=2 \rho \cos \varphi$, $E_1''=(d^2E/dq^2)_{q_1}=-8i \kappa+ 2 \rho \sin \varphi$, and $E_2''=(d^2E/dq^2)_{q_2}=-8i \kappa - 2 \rho \sin \varphi$.
 For large $t$, the functions under the sign of integrals on the right hand sides of Eq.(10) are narrow at around $\xi=0$, so that the integrals can be extended from $-\infty$ to $\infty$, finally yielding
  \begin{eqnarray}
  c_n(t) & \sim  & \sqrt{\frac{1} {2 \pi i E_1" t}} \exp(iq_1 n-iE_1 t) \exp\left[- \frac{(n-E_1't)^2}{2iE_1''t} \right]  \nonumber \\
  & + &  \sqrt{\frac{1} {2 \pi i E_2" t}} \exp(iq_1 n-iE_2 t) \exp\left[- \frac{(n-E_2't)^2}{2iE_2''t} \right] \;\;\;\;\;\;\;
  \end{eqnarray}
     \begin{figure}[htbp]
  \includegraphics[width=86mm]{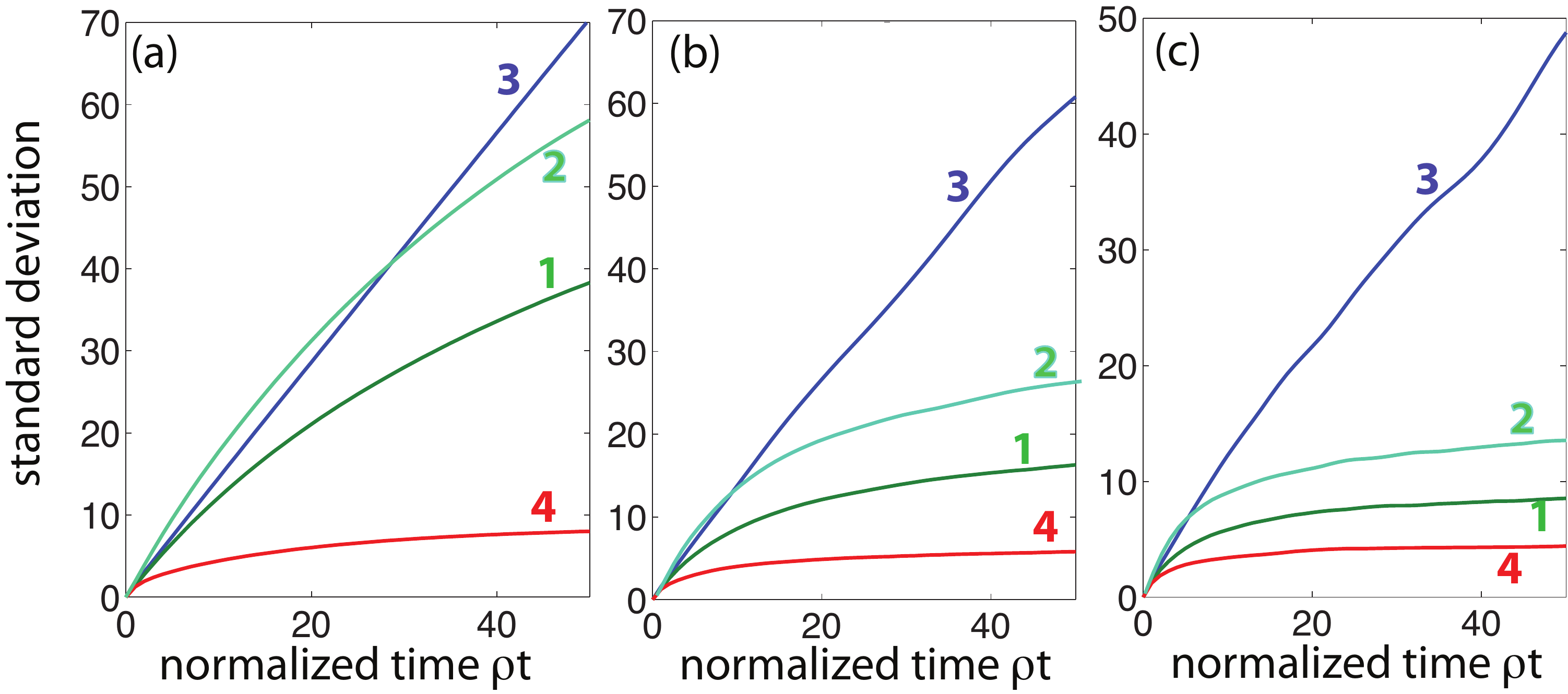}\\
   \caption{(color online) Temporal behavior of the standard deviation of the probability distribution $|p_n(t)|^2$, for initial single-site excitation, in a disordered lattice with $U_n$ random variable uniformly distributed in the interval $(-\Delta,\Delta)$: (a) $\Delta=0.5$, (b) $\Delta=1$, (c) $\Delta=1.5$. Curve 1 refers to the Hermitian lattice ($\kappa=\gamma=0$, $\rho=1$, $\varphi=0$), curves 2, 3 and 4 refer to non-Hermitian lattices ($\kappa=0.3$, $\gamma=2 \kappa=0.6$, $\rho=1$) with flux values $\varphi=0$, $\varphi= \pi/4$ and $\varphi= \pi/2$, respectively. The curves are obtained after averaging over 100 realizations of disorder.}
\end{figure}
 Equation (11) clearly shows that the asymptotic form of $c_n(t)$ is given by the superposition of two damped and dispersive Gaussian wave packets, which propagate at opposite directions with the group velocities $v_{g1}=E_1'=- 2 \rho \cos \varphi$ and $v_{g2}=E_2'= 2 \rho \cos \varphi$. Note that at $\varphi= \pi/2$ the group velocities vanish, there is not transport (propagation) along the lattice and the two dispersive Gaussian wave packets at rest interference and undergo a diffusion broadening, similar to the case studied in Ref. \cite{r30}; see Fig.3(c). On the other hand, for $\varphi=0$ the group velocities are largest and two oppositely-propagating Gaussian wave packets are clearly observed; see Fig.3(a). For $\varphi= \pi/4$, a similar scenario is found, but with a slower group velocity [Fig.3(b)].\par
 Let us now consider the case of a disordered lattice by assuming $U_n$ a random variable with uniform distribution in the range $(-\Delta,\Delta)$ (diagonal disorder). While for the ordered lattice the spreading dynamics for the $\varphi=0$ and $\varphi= \pi/4$ are similar [see Figs.3(a) and (b)], a deeply different behavior is found when considering lattice disorder. In fact, while in the $\varphi=0$ case disorder prevents transport via Anderson localization like in the Hermitian limit $\kappa=0$, non-Hermitian delocalization is observed in the $\varphi= \pi/4$ case even for relatively large values of lattice disorder, as shown in Figs.4 and 5. The reason thereof is that for $\varphi=0$ one has $E_1=E_2$, so that back reflections arising from elastic scattering at lattice defects or disorder are allowed like in an ordinary Hermitian lattice, resulting in Anderson localization. On the other hand, for $\varphi= \pi/4$ one has $E_1 \neq E_2$: backscattering is therefore prevented because reflected waves are {\it evanescent} rather than propagative ones \cite{r38}, and non-Hermitian delocalization is correspondingly observed. In the following we will therefore limit to consider the $\varphi= \pi/4$ case. Nevertheless,  it is worth to briefly discuss the impact of the flux $\varphi$ on Anderson localization in the non-Hermitian cases $\varphi=0, \pi/2$, as observed from the numerical results shown in Fig.5. The figure clearly shows that, as compared to the Hermitian lattice, localization is weaker in the non-Hermitian lattice with zero flux $\varphi=0$, whereas it is stronger for the flux $\varphi= \pi/2$. Such a result can be explained after observing that in the Hermitian  one-dimensional Anderson model localization turns out to be relatively weak at the band center and comparatively strong near the band edges \cite{r56}. As the non-Hermitian next-nearest neighbor hopping is introduced, the imaginary part of the energy dispersion curve acts as a filter in the reciprocal (wave number) space and, as discussed above, the dominant modes in the dynamics are those with Bloch wave number $q$ near $q=q_{1,2}= \pm \pi/2$ (see Fig.2). In the $\varphi=0$ case, such modes are center band modes with lower degree of localization, which explains why Anderson localization is less effective for $\varphi=0$ as compared to the Hermitian lattice. On the other hand, for $\varphi=\pi/2$ the filtered modes are the band edge modes which show stronger localization. 
   \begin{figure}[htbp]
  \includegraphics[width=86mm]{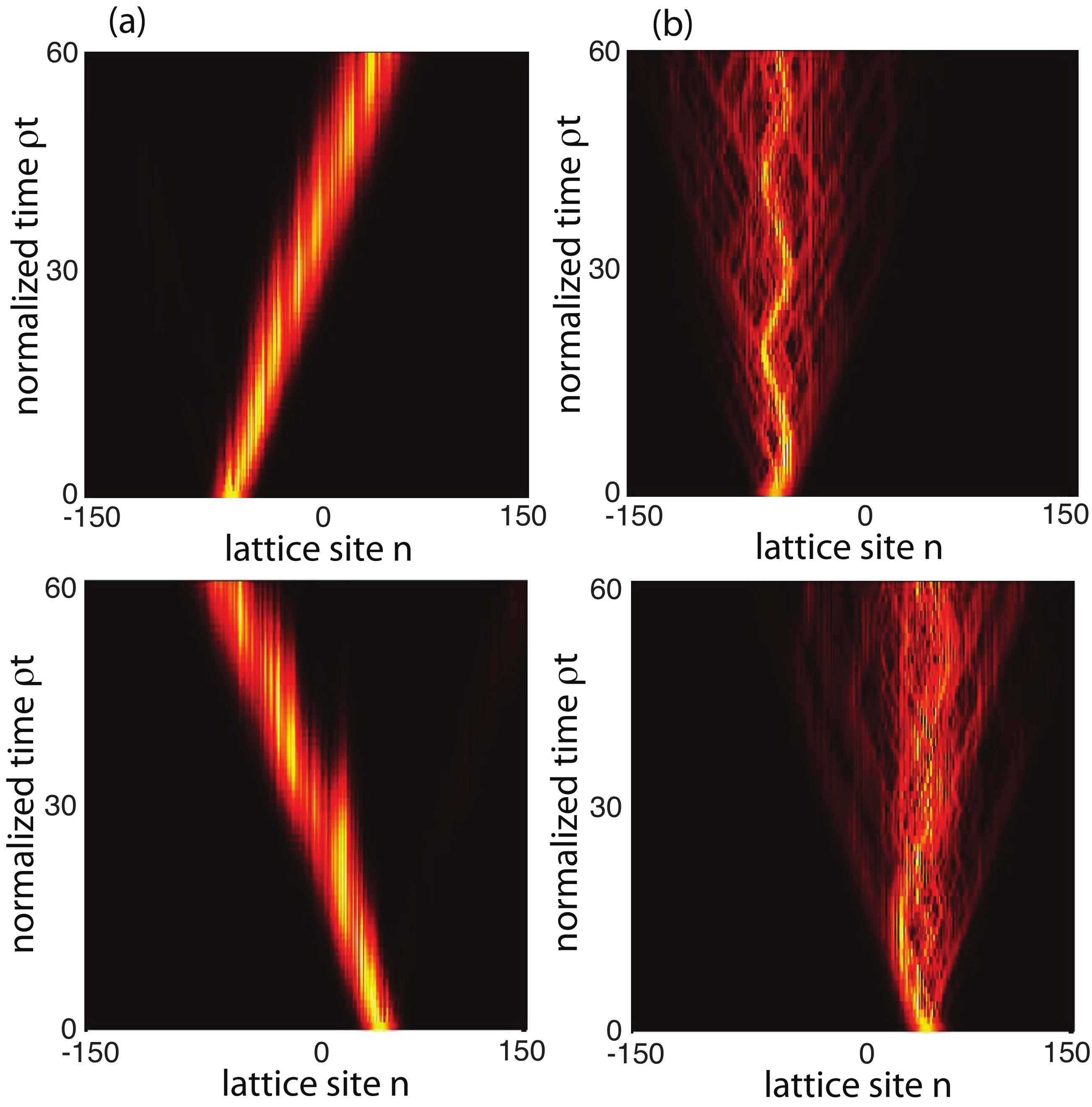}\\
   \caption{(color online) Propagation of a Gaussian wave packet (snapshot of the normalized amplitude probability $|p_n(t)|$ in a pseudo color map) in a disordered lattice with $U_n$ random variable uniformly distributed in the interval $(-\Delta,\Delta)$, with $\Delta=1$. (a) Non-Hermitian lattice ($\kappa=0.3$, $\gamma=2 \kappa=0.6$, $\rho=1$, $\varphi=\pi/4$);  (b) Hermitian lattice ($\kappa=\gamma=0$, $\rho=1$, $\varphi=0$). The upper panels refer to a forward-propagating wave packet (carrier wave number $q_0= -\pi/2$, width $w_0=10$), whereas the lower panels refer to a backward-propagating wave packet (carrier wave number $q_0= \pi/2$, width $w_0=10$).}
\end{figure}

   \begin{figure}[htbp]
  \includegraphics[width=86mm]{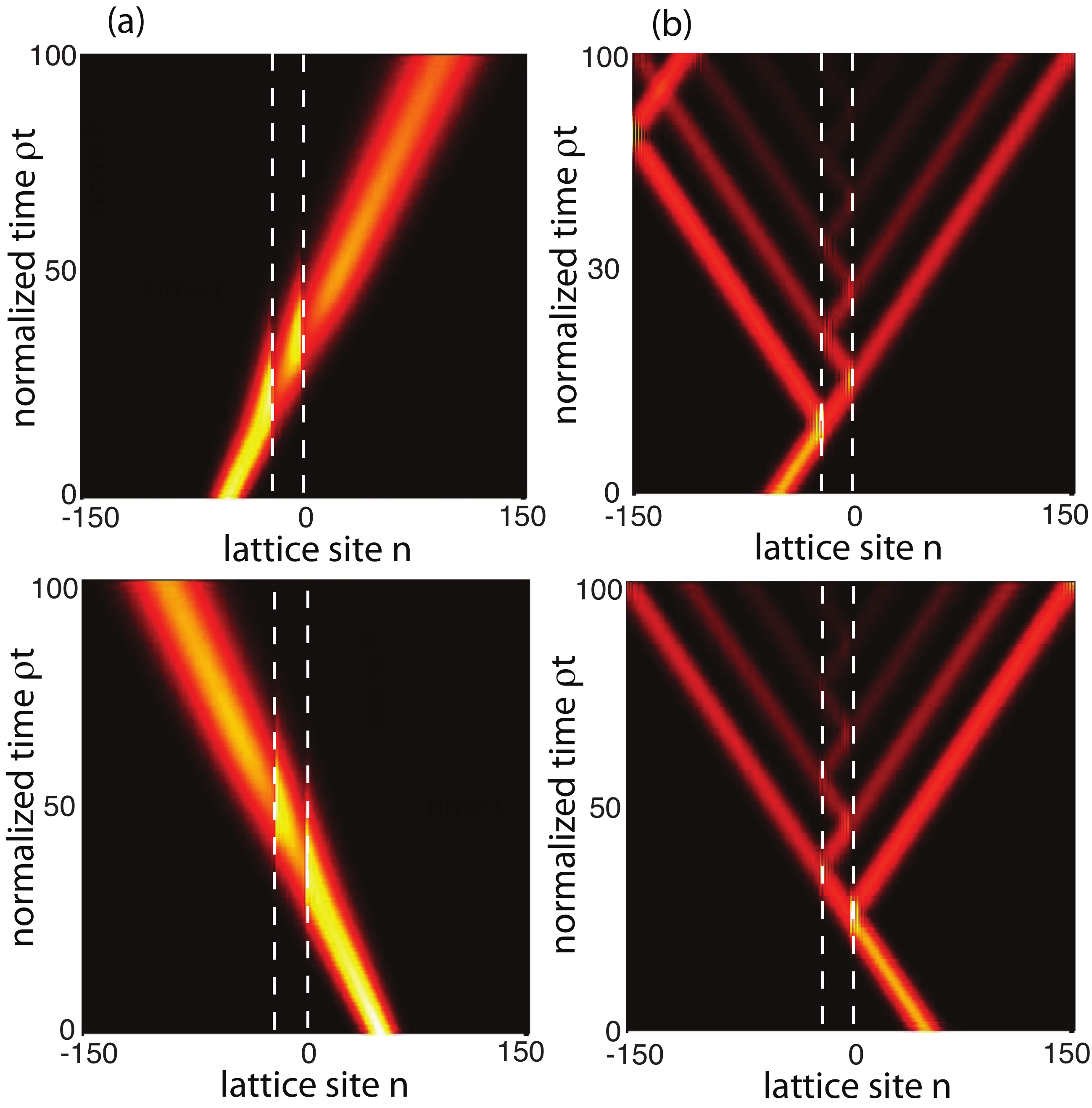}\\
   \caption{(color online) Same as Fig.6, but in an ordered lattice with two defects at sites $N_1=-20$ and $N_2=0$ ($V_0=1$, $N_2-N_1=20$). Other parameter values are as in Fig.6. The vertical dashed lines show the position of the two defects.}
\end{figure}
  
  \section{Bidirectional robust transport}
  The results shown in the previous section suggest that robust and {\em bidirectional} transport of wave packets, which is immune to lattice disorder or structural defects, should be observable in a non-Hermitian lattice with a flux $\varphi= \pi/4$ or around such a value, where the group velocity does not vanish and $E_1 \neq E_2$. Numerical results confirm such a prediction when wave packet propagation is simulated in a lattice with either on-site energy potential disorder or structural lattice imperfections. 
  
  As an example, Fig.6 shows the numerically-computed evolution in a disordered lattice of an initial Gaussian wave packet 
  
  \begin{equation}
  c_n(0) \propto \exp[-(n-n_0)^2/w_0^2+iq_0n]
  \end{equation}
  with carrier wave number $q_0=\pm \pi/2$, spatial width $w_0$ and initial position $n_0$, that propagates either forward or backward along the lattice with group velocity $E'_{1,2}= \mp \sqrt{2} \rho$. The results are plotted for either an Hermitian lattice [Fig.6(b)] and a non-Hermitian lattice with flux $\varphi= \pi/4$ [Fig.6(a)]. In the former case disorder clearly yields back reflections and prevents propagation of the wave packet along the lattice, regardless of the propagation direction. This behavior arises because of Anderson localization. Conversely, in the non-Hermitian lattice propagation of the wave packet is preserved for both propagation directions in spite of disorder.

  Robust bidirectional transport is observed also when the lattice shows structural defects rather than disorder. Let us consider, as an example, a lattice with two potential defects at sites $N_1$ and $N_2$, i.e. let us assume in Eq. (6) $U_n=V_0(\delta_{n,N_1}+\delta_{n,N_2})$, where $V_0$ is the strength of the potential
defect. In the Hermitian lattice a propagating wave packet undergoes multiple reflections back
and forth between the two defects, like in a Fabry-Perot cavity. This yields multiple transmitted waves (echoes), as shown in Fig.7(b). Conversely, in the non-Hermitian lattice multiple reflections are prevented and a single transmitted wave, without any reflected wave packet, is observed for both left and right incidence sides [Fig.7(a)].
  
\section{Conclusions}

The realization of robust transport, which is immune to backscattering and disorder, is of major importance in different areas of physics. In this context, a huge attention has been devoted to the study of topologically-protected edge states, which can propagate unidirectionally at the
boundaries of two or three-dimensional systems.  Such states are found in a wide
variety of quantum Hall systems, topological insulators, and topological superconductors. Recently, it has been suggested that unidirectional robust transport, which does not rely on topological protection,  can arise in non-Hermitian one-dimensional lattices as a result of lattice band engineering \cite{r37}, a paradigmatic example being provided by non-Hermitian transport in the Hatano-Nelson model \cite{r3}. In this model the complex energy spectrum induced by an imaginary  vectorial field forbids wave propagation in one direction of the lattice, thus favoring unidirectional transport which is immune to structural lattice imperfections and disorder. While unidirectional robust transport is a well established property, a major open question is whether one can realize {\it bidirectional} robust transport. In the present work we have shown that, by exploiting robustness of non-Hermitian transport, bidirectional transport, which is immune to disorder and backscattering, can be realized in a quasi one-dimensional (zigzag) lattice with imaginary (non-Hermitian) hopping amplitude and a synthetic gauge field. By properly tuning the gauge field, non-Hermitian delocalization and persistence of transport despite disorder or structural imperfections is observed in {\it both} directions. The implications of such a result are twofold. On the one hand, it sheds new light into the properties of non-Hermitian transport, providing a new possible route toward the realization of robust transport in quasi-one dimensional tight-binding lattice systems that does not rely on topological protection. On the other hand, while robust transport in topologically-protected systems is based on chiral symmetry breaking, i.e. it is a directional (one-way) robust transport, the suggested non-Hermitian transport in a zigzag system provides a rather unique example of {\em bidirectional} robust transport. Such a distinctive feature dispels the rather general belief that that robust transport is a directional effect, thus opening up new routes toward the engineering of quantum or classical transport exploiting non-Hermitian dynamics.



\end{document}